\documentclass{article}

\usepackage{arxiv}

\usepackage[utf8]{inputenc}
\usepackage[T1]{fontenc}
\usepackage{hyperref}
\usepackage{url}
\usepackage{booktabs}
\usepackage{amsmath}
\usepackage{amsfonts}
\usepackage{nicefrac}
\usepackage{microtype}
\usepackage{cleveref}
\usepackage{lipsum}
\usepackage{graphicx}
\usepackage{natbib}
\usepackage{doi}

\usepackage{algorithm}
\usepackage{algpseudocode}

\usepackage{listings}
\usepackage{csquotes}
\usepackage{comment}
\usepackage[title]{appendix}
\usepackage{tabularx}
\newcolumntype{L}{>{\raggedright\arraybackslash}X}

\usepackage{xcolor}

\title{Towards Adaptive and Real-Survey-Aligned Simulated Urban Mobility Behaviour of Generative Agents}

\date{}

\author{\href{https://orcid.org/0009-0003-8659-7476}{\includegraphics[scale=0.06]{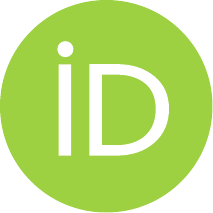}\hspace{1mm}Yves Bru}\thanks{Corresponding author} \\
	UMR IRIT\\
	Université Toulouse Capitole\\
	Toulouse, France \\
	\texttt{bru.yves@protonmail.com} \\
	\And
	\href{https://orcid.org/0000-0002-0334-3055}{\includegraphics[scale=0.06]{orcid.pdf}\hspace{1mm}Kamaldeep Singh Oberoi} \\
	CESI\\
	CESI LINEACT\\
	Toulouse, France \\
	\texttt{ksoberoi@cesi.fr} \\
    \And
    \href{https://orcid.org/0000-0002-9005-3004}{\includegraphics[scale=0.06]{orcid.pdf}\hspace{1mm}Benoit Gaudou} \\
	UMR IRIT\\
	Université Toulouse Capitole\\
	Toulouse, France \\
	\texttt{benoit.gaudou@ut-capitole.fr}
}

\renewcommand{\shorttitle}{Towards Adaptive and Real-Survey-Aligned Simulated Urban Mobility Behaviour of Generative Agents}

\begin{document}
\maketitle
\begin{abstract}
In existing agent-based simulations of multimodal urban mobility, agents, representing humans, choose their travel modes either using discrete choice or machine learning models fitted on structured travel survey data, or with rules explicitly written by the domain experts. Such approaches fix \emph{a priori} the agents' possible behaviour space. Hence, a factor absent from variables or rules cannot affect agents' decisions. In reality, choosing a transport mode is a multidimensional decision in which objective and subjective factors intertwine and change over time. Integrating such complexity into a city-scale multimodal mobility simulation remains a challenge. Generative agents (GAs) built on large language models promise to respond to this challenge because they carry decision heuristics that travel surveys do not record. However, to the best of our knowledge, aligning their decisions with real household travel surveys remains an open question. In this paper, we build an agent-based model of daily multimodal mobility for Toulouse (France). We compare the performance of fifteen decision-makers (simple baselines, models fitted on the survey, GAs with minimal and tuned prompts, and a typed classifier) on the real household travel survey. Our results show that GAs with tuned prompts come close to the fitted models on the modal split of a simulated day without transport disruptions, but not trip by trip. GAs also react when unforeseen events occur. The recently released Jev-1.13 TypeSafe classifier gives surprisingly great results, showing that generating output text is not necessary. The results presented in this paper would be of interest to transport policy makers and service providers looking to apply generative agent-based modelling for improving urban mobility services.
\end{abstract}

\section{Introduction}
\label{sec:introduction}

Agents in existing multi-agent simulations of multimodal urban mobility choose modes either using discrete choice or machine learning models on structured data \citep{train2009discrete,martinbaos2023prediction}, or with explicit expert-defined rules \citep{smith1995transims,horni2016matsim}. The models are fitted on household travel surveys, where a territory's residents describe every trip they made on one day. In both approaches, the space of agents' representable behaviours is fixed \emph{a priori}: a factor encoded as neither variable nor rule cannot influence any decision. This limits the adaptability of such agents and models over time and to unforeseen situations.

Generative agents (GAs) built on Large Language Models (LLMs) promise to lift this limit. Trained on vast amounts of text, they carry decision heuristics absent from surveys and adapt without explicit rules. They also appear to possess partial knowledge of human behaviour, allowing them to simulate human-like choices \citep{suri2024heuristics}. If these choices proved realistic, GAs would open scenarios that exist only in text. A city could, for instance, compare two wordings of a public campaign, test the effect of press coverage of safety incidents in the metro, or anticipate the reaction to an employer's telework announcement.

How close the mode choices made by GAs come to mode choices that humans make, however, has seldom been tested in the literature. Most published evaluations judge a GA against a rule-based agent, or against aggregate patterns generated synthetically. Using GAs to study real scenarios is worth their cost if their promise holds. In this paper, we argue that this promise rests on five assumptions, which we test one-by-one against a household travel survey and paired experiments:

\begin{enumerate}
  \item[(A1)] GAs bring built-in, albeit partial and approximate, knowledge of people's travel behaviours to their choices;
  \item[(A2)] Their behaviour can be adjusted by tuning the prompt with criteria stated using text;
  \item[(A3)] Their global mode share performance is close and comparable to models fitted on a travel survey;
  \item[(A4)] Their choices are also realistic for each individual;
  \item[(A5)] They can react to \emph{unforeseen events} provided as a textual scenario, with no new variable or re-estimated model.
\end{enumerate}

Our objective is to determine whether, and where, text-based models are worth their cost \citep{luccioni2024power} as the decision module of GAs in terms of mode choice. We compare 15 decision-makers: simple baselines without any behavioural knowledge (uniform draw, all-car, fastest option), reference models (such as the standard discrete choice model) and classical Machine Learning classifiers trained on the survey's trips, as well as text-based models (3 LLMs and the recently released typed classifier \texttt{Jev-1.13 TypeSafe model} \citep{typesafe2026jev}), first run without prompt engineering, then tuned by editing their prompt.

Our case study covers the 453 municipalities around Toulouse (France) included in the 2023 household travel survey \citep{tisseo2023emc2}. We score the 3,299 trips that 1,000 synthetic \emph{personas}, checked against the survey, make on one day. Each persona describes a person's attributes, household, and planned trips, and runs as an \emph{agent} in a simulation of the study area with its actual networks and timetables. The simulation extends an existing architecture \citep{vu2025modeling} that couples the agent-based simulation platform GAMA \citep{taillandier2019gama} with the OpenTripPlanner router \citep{opentripplanner2025} and an LLM.

This paper makes the following three contributions, which together test the five above-mentioned assumptions:

\begin{enumerate}
  \item \emph{Built-in knowledge and prompt tuning} (A1 and A2). We show that, using only a minimal prompt, text-based models close between 50 and 85\,\% of the distance from the best baseline to the best reference model. Prompt tuning brings the best of them close to the performance of the reference models, and Jev within their range.
  \item \emph{Evaluating GAs on aggregated mobility and individual trip scales (A3 and A4).} We
  show that the aggregated modal split of the best GAs comes close to that of the reference
  models, and falls within their range for Jev (tuned prompt). Trip by trip, however, every GA
  gives less probability to the reported mode than any reference model, the most accurate one
  only narrowly. A close aggregated modal split therefore does not establish realistic
  individual choices.
  \item \emph{Sensitivity to an unforeseen event} (A5). We show, in a text-described scenario, a
  suspicious noise in the car engine, that the exposed agent leaves the car for sixteen days, a
  response no model restricted to survey variables can produce.
\end{enumerate}

The paper is organised as follows. Section \ref{sec:related} reviews the related work and positions our contributions in the literature. Section \ref{sec:architecture} presents the software architecture used in our paper. Section \ref{sec:results_ordinary_day} presents the results obtained and tests the assumptions A1 to A4 against the household survey on an ordinary day. Section \ref{sec:unforeseen_events} tests the assumption A5 on an unforeseen event. Section \ref{sec:conclusion} concludes the paper.

\section{Related work}
\label{sec:related}

\subsection{Mobility mode choice}
For more than half a century, discrete choice models have predicted travel mode choice. Such a model computes each mode's probability from travel time, cost, and the traveller's attributes \citep{mcfadden1974conditional,benakiva1985discrete}. Once fitted on a travel survey, it reproduces the survey's mode shares through each mode's alternative-specific constant \citep{train2009discrete}. However, the weights given to time and cost, and their spread across travellers, never change after estimation. Agent-based simulators that delegate mode choice to such a model inherit this rigidity. For example, the eqasim framework couples the MATSim simulator \citep{horni2016matsim} with discrete mode choice models \citep{horl2018bridging,horl2021pipeline}. Testing a new transport policy in such scenarios therefore calls for re-estimation of the weights or entirely new costly surveys \citep{lammer2026gta}. Behavioural research documents factors that surveys do not record, such as habit and perception. Travellers with a strong mode habit gather less information about the alternatives and use simpler decision rules \citep{verplanken1997habit}. Experience also filters perception. For example, in \citep{adam2025survey}, among 650 survey respondents, car users rated the car more affordable than non-users did. Our reference models cannot represent such filters, and GAs have been proposed to fill that gap.

\subsection{Generative agents in mobility simulation}
GAs now simulate the mobility of whole cities, but few studies compare their mode choices with the observed ones. Generative agents pair an LLM with a memory of their experiences and a reflection step \citep{park2023generative}. This design has been applied to travel demand modelling \citep{liu2024toward}. CitySim \citep{bougie2025citysim} and its successor CityReal \citep{bougie2026cityreal} scale such agents to a city. Only CityReal compares the simulated modal split with an observed one. However, its survey data are proprietary, and it runs no reference model on mode choice. GTA's \citep{lammer2026gta} agents only choose their modes, and an existing simulator (SUMO) moves them through the network. However, each GTA agent returns one mode per trip rather than a distribution, so its scores cannot be weighted by probability. Its only point of comparison is a naive predictor, the observed split of another German state, and that predictor wins.

On the platform side, some work couples generic agent-based modeling and simulation platforms, such as the GAMA platform \citep{taillandier2019gama} with GAs. As an example, \citep{vu2025modeling} simulates five agents, which GAs choose transit itineraries in Toulouse over one simulated month, with a daily reflection feeding a long-term memory. The study measured how fast these choices stabilised into habits rather than how well they match the observed ones. Other studies expose GAs to more than an ordinary day. For example \citep{alves2026evaluating} couples GAMA platform to an LLM module with a persistent memory, which decides whether an agent re-plans after a disruption. It compares this module with a rule-based baseline. Two further studies compare such agents with observed behaviour, though not on mode choice. In the first, agents learn their route over successive days, and their choices are compared with a route-choice experiment \citep{liu2025aligning}. In the second, agents live through the July 2024 Philadelphia heatwave, and their activities are compared with a survey collected during the heatwave \citep{xia2026empirical}.

\subsection{Distributional alignment of LLM populations}
There has been a recent growing interest in asking whether LLM populations can stand in for human samples; it shapes how such agents are scored. For example, in \citep{argyle2023outofone}, conditioning a model on sociodemographic attributes recovered the matching subpopulation's response patterns. A model aligns better when it states a distribution in words than through its token probabilities or sampled answers \citep{meister2024benchmarking}.

\subsection{Our positioning}
None of the works above combines the three elements of our contributions: (i) a deep comparison of GAs' performance \emph{w.r.t.} reference models and prompt tuning to enhance these results, (ii) a comparison with these reference models at two scales (aggregated mobility and individual trip), and (iii) a paired test of an event that no survey records. These three points are the focus of this paper and make our contribution original.

\section{Software architecture}
\label{sec:architecture}

The system simulates the daily travel of Toulouse-area residents. Each day, every resident carries out their planned activities and, for each trip, chooses an itinerary among those that the actual road, cycling, and public transport networks offer at that time. Each agent runs one persona (resident's description) of the population. It also has a location in the simulated world, a memory, routing engines that compute each trip's candidate itineraries, and a decision-maker to make choices. The decision-maker can be a prompted LLM, a typed classifier like Jev-1.13, a model estimated on the survey's trips, or a simple baseline. Each agent plans its trip, waits for the departure time, then follows the chosen itinerary on foot, by bicycle, or by car. When using bus, metro, or train, it waits at designated stops, and each vehicle carries it onward. GAMA moves the agents every two simulated minutes. This section describes how a trip is decided and carried out, the information the decision-maker reads and the distribution it produces, and the agent's memory.

\subsection{How a trip is decided and carried out}
\label{sec:loop}

\begin{figure}
  \centering
  \includegraphics[width=\textwidth]{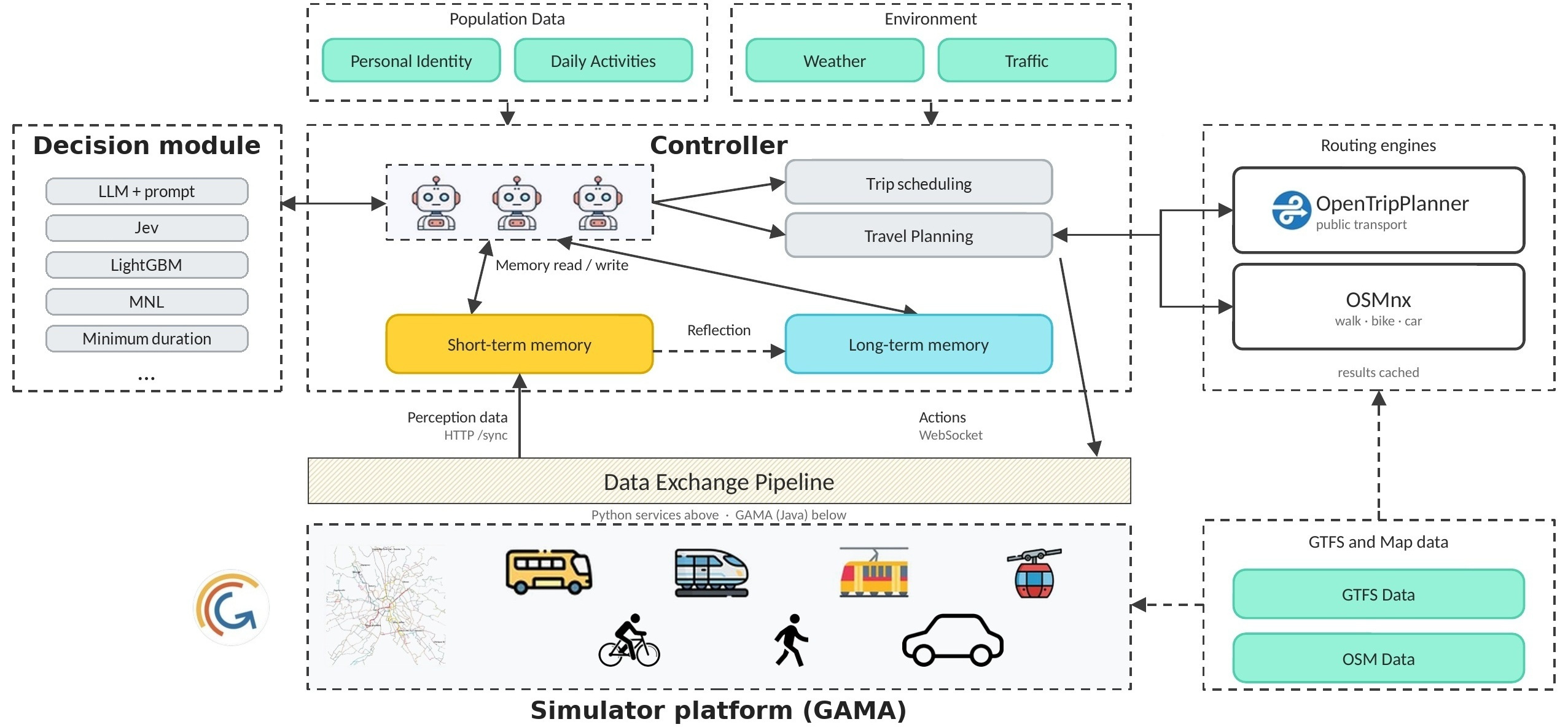}
  \caption{System architecture: the simulator platform (bottom), the controller (centre), and the decision module (left).}
  \label{fig:architecture}
\end{figure}

The architecture has three components (Figure \ref{fig:architecture}):
\begin{itemize}
  \item \textbf{A simulator platform} (bottom) holds its physical state: a position, a daily schedule and the trips it physically makes. The world runs on GAMA, with the real geography of the study area, its networks and the clock.
  \item \textbf{A controller} (centre) runs the agent lifecycle and holds its state: its short-term and long-term memory and the current location of each household vehicle.
  \item \textbf{A decision module} (left) hosts the decision-maker.
\end{itemize}

At departure time for each agent, the controller requests the itineraries from the routing engines and builds the trip description. The decision-maker chooses among the available itineraries. The simulation carries out the trip and reports what happened (transfers, waits, arrival time). This feedback enters the memory that informs the next decision.

Formally, a trip is defined as a tuple $T = (p, O, D, A, t)$, where agent $p$ travels from origin $O$ to destination $D$ for activity $A$, starting at time $t$. An itinerary for $T$ is a sequence of legs $i = (\ell_1, \dots, \ell_K)$ leading from $O$ to $D$. Each leg uses a single mode and a single line for public transport. Let $\mathcal{I}(T)$ be the set of itineraries offered for $T$. A decision-maker maps a trip $T$ to a probability distribution over $\mathcal{I}(T)$. The two routing engines cover different networks. OpenTripPlanner produces the public transport itineraries from actual timetables. A fastest-path search with OSMnx \citep{boeing2025osmnx} on the OpenStreetMap graph produces itineraries on foot, by bicycle and by car. Each edge is weighted by its free-flow travel time, at a speed set per mode and road class. Signal delays are then added along the path and, for the car, each edge time is scaled by hourly congestion. $\mathcal{I}(T)$ contains only the itineraries that respect the distance limits and the vehicles actually available where the agent is. Every decision-maker chooses from this set. In case of a GA, the LLM is called for four tasks. It makes the decision, rates an event as it enters the memory, consolidates the memory in the evening and, every few days, re-reads the complete long-term memory as a self-reflection that updates what the agent holds true of itself.

As in \citep{vu2025modeling}, GAMA is coupled to a separate Python server hosting the decision-makers and, in particular, GAs, over two channels. HTTP and WebSocket protocols handle the synchronous and asynchronous communication between GAMA platform and the Python Server. This design suits our work for two reasons. First, since the decision-maker runs outside the simulation, it can be replaced without changing the world or the routing. Second, the end-of-day reflection revisits the agent's existing beliefs rather than starting afresh, so the agent's experience is consolidated day-by-day and habits can form. This paper introduces some changes in the architecture of \cite{vu2025modeling}, including GA decisions as a probability distribution, and the possibility to replace GAs with any model for decision-making (see \href{https://hal.science/hal-05797773}{Supplementary Material}: Appendix B.5 for more details).

\subsection{The decision prompt and its answer}
\label{sec:receives}

The agent numbers the itineraries of $\mathcal{I}(T)$ (Section \ref{sec:loop}) in random order, so that on average no option gains from appearing first. It places them in a trip description containing:

\begin{itemize}
  \item \textbf{The traveller.} The persona summary, with age, occupation, and the household's vehicles and public transport passes.
  \item \textbf{The trip.} Destination, departure time, and the weather now and later in the day.
  \item \textbf{The rest of the day.} The trips left before returning home.
  \item \textbf{Past experience.} The memories retrieved for this trip, none in a benchmark run (Section \ref{sec:protocol}).
\end{itemize}

The agent then delegates the choice to its decision-maker. LLM assigns each option a probability and writes these numbers in a structured JSON object. These probabilities are part of the text the model writes, like the reason that follows them.

\newcommand{\jl}[1]{\par\noindent\hbox{\ttfamily #1}}
\definecolor{jkey}{RGB}{0,60,140}
\definecolor{jstr}{RGB}{90,90,90}
\newcommand{\jk}[1]{\textcolor{jkey}{#1}}
\newcommand{\js}[1]{\textcolor{jstr}{#1}}
\newcommand{\jn}[1]{\textbf{#1}}
\newcommand{\ji}[1]{#1}

\begin{center}
\begin{minipage}{0.65\textwidth}
\begingroup\parskip=0pt\topsep=0pt
\jl{\{}
\jl{~\jk{"probabilities"}: [}
\jl{~~\{\jk{"index"}: \ji{0}, \jk{"mode"}: \js{"foot"}, \jk{"prob"}: \jn{5}\},}
\jl{~~\{\jk{"index"}: \ji{1}, \jk{"mode"}: \js{"foot,bus,foot"}, \jk{"prob"}: \jn{30}\},}
\jl{~~\{\jk{"index"}: \ji{2}, \jk{"mode"}: \js{"foot,metro,foot"}, \jk{"prob"}: \jn{15}\},}
\jl{~~\{\jk{"index"}: \ji{3}, \jk{"mode"}: \js{"car"}, \jk{"prob"}: \jn{50}\}}
\jl{~],}
\jl{~\jk{"reason"}: \js{"Rain; the car avoids a long walk."}}
\jl{\}}\par\endgroup
\end{minipage}
\end{center}

We call this distribution the probability mass. The agent draws the decision at random from it, rather than keeping the most probable option: in this example, the person takes the car in half of the draws. The reference models also yield a probability for each option, from which the agent draws in the same way. They read variables on the person, the household, and the trip instead of text.

\subsection{Short-term and long-term memory}
\label{sec:memory}

Two memories hold what an agent has been through. A short-term memory receives the decisions and the physical experience the simulation reports. Each evening, a consolidation pass writes from it into the long-term memory. This pass adapts the reflection of generative agents \citep{park2023generative}, which reflect once recent events pass an importance threshold; ours reflect at a fixed time each day. In our system, each household member also tells the others about their day in the evening, and that account enters the listeners' reflection.

The long-term memory holds two kinds of entries, which map onto the episodic and semantic memories distinguished in language agents \citep{sumers2024coala}. An episodic trace records what happened and when. A belief states what the agent holds true of the world or of itself, and its confidence rises or falls with new observations. Memory research has long held that episodic memory is more prone to forgetting than semantic memory \citep{renoult2020historical}, so we let only traces decay.

Each trace carries a severity $g \in [0, 1]$, which measures how much the event disrupted the agent's day. For a trip, $g$ is computed from the delay, missed connections, and incidents the simulation measured. An unforeseen event is a short first-person text written into the short-term memory when the trip it disrupts ends, with that trip's record, and scored the same way. Only a text the agent reads is rated by the LLM. Severity lets a breakdown outweigh a routine trip in memory. It acts twice: a severe trace ranks higher and lasts longer. The first role matches the importance score of generative agents \citep{park2023generative}, whose recency decays at the same rate for every memory.

A trace's weight falls exponentially, as in MemoryBank \citep{zhong2024memorybank}: a routine trip is mostly forgotten within a week, while a severe event still weighs weeks later. The past-experience part presents the past in four blocks: the agent's habits (what it does most often), its beliefs, its recent events, ranked by severity, and three memories retrieved for the trip at hand. Retrieval keeps the candidates closest to the trip's description, about a mode offered now, or among the most severe, and scores each between 0 and 1 as a weighted sum of five criteria: similarity to the current trip, recency after decay, severity, match on context (mode, time slot, purpose or place), and weather. The weights are set by hand, not fitted, and similarity weighs the most (\href{https://hal.science/hal-05797773}{Supplementary Material}: Appendix E).

\section{Results on the ordinary day}
\label{sec:results_ordinary_day}

In this section, we present the results obtained on an ordinary day and discuss four of the five assumptions (Section \ref{sec:introduction}) that the survey describes.

\subsection{Experimental protocol}
\label{sec:protocol}

\paragraph{Population.} 
Every score in this section comes from one evaluation population of 1,000 synthetic personas (499 complete households of the study area), generated with the eqasim synthesis pipeline \citep{horl2021eqasim} from census and national travel survey data. A disjoint training population, drawn the same way from the rest of the study area, serves prompt tuning only. The evaluation population is aligned against the local survey. The survey \citep{tisseo2023emc2} follows the national protocol of French household travel surveys (details and comparisons in \href{https://hal.science/hal-05797773}{Supplementary Material}: Appendix A). It drew 10,783 households at random, interviewed between September 2022 and February 2023 outside school holidays. The 15,775 respondents describe each trip of the previous weekday: origin, destination, purpose, times and modes used. Weights then align the sample with the census on household size, car ownership, age and occupation. On thirteen marginal distributions (e.g., age class, car ownership), no category of the population departs from its survey share by more than 0.50 points, but it travels slightly less: 3.30 trips per persona against 3.53. Agents start with an empty memory, so no score depends on the memory constants of Section \ref{sec:memory}.

\paragraph{Decision-makers.}
Fifteen decision-makers choose among same options on every trip of the population: 

\begin{itemize}
    \item[(i)] Three \textbf{baselines} with no behavioural knowledge (a uniform draw over the options offered, all-car, and the fastest option (minimum duration);
    \item[(ii)] Four \textbf{reference models} trained on the survey's trips (the multinomial logit (MNL), the standard discrete choice model, LightGBM gradient boosting \citep{ke2017lightgbm}, Kernel Logistic Regression (KLR) \citep{zhu2005kernel} and Random Forest (RF) \citep{breiman2001random});
    \item[(iii)] Three \textbf{LLMs} (Gemini 3.1 Flash-Lite, Gemini 3.5 Flash-Lite and Mistral Large), each run under the minimal prompt and under its own tuned prompt leading to six distinct decision-makers);
    \item[(iv)] \textbf{Jev} (\texttt{Jev-1.13 TypeSafe model}, version \texttt{jev-1.13.0}), runs under the minimal prompt and tuned prompt given two distinct decision-makers. Like the LLMs, it reads the trip description and returns a probability for each option offered.
\end{itemize}

Each text-based decision-maker thus combines a model and a prompt. The minimal prompt gives the model the facts of the trip and the output format, with no general criteria. The tuned prompt keeps that format and adds general criteria stated in words, tuned on the training population, separately for each model (Section \ref{sec:tuning}). Below, Gemini 3.1 and Gemini 3.5 stand for the two Flash-Lite models; figures and tables give the prompt in brackets, as in Gemini 3.5 (tuned). Every decision-maker except the baselines reads the same 21 core variables on the person, the household, and the trip. Beyond this core, both kinds of model receive transport supply, the weather and the rest of the day, in different forms. The text-based decision-makers read the itineraries as text, leg by leg. The reference models read fifteen numbers, 36 variables in all (\href{https://hal.science/hal-05797773}{Supplementary Material}: Table 0.3), computed between the centres of the origin and destination zones in five time bands, because the survey locates trips only by zone. The four reference models were estimated under identical conditions, on the same 39,203 survey trips, which no text-based decision-maker saw. The prompts saw the survey only through aggregate gaps on the training population during tuning.

\paragraph{Comparison of models.}
\label{sec:rules}
Two rules guard against two ways of biasing a comparison between a GA and a reference model:

\begin{itemize}
    \item Rule 1 scores the probability mass a decision-maker puts on each option, not the option it ranks first, which the agent does not necessarily take (Section \ref{sec:receives}).
    \item Rule 2 handles a difference in output. A reference model predicts a share for every mode, whereas a text-based decision-maker sees only the options offered. We therefore restrict each prediction of a reference model to the modes offered for that trip, then rescale it to 100\,\%. When several options share a mode, that mode's share is split evenly among them.
\end{itemize}

\paragraph{Indicators.}
\label{sec:quantities}

Indicators are computed at two scales: the modal split of the population and the agreement on each trip (Section \ref{sec:individual}). The modes are car, public transport, walking and cycling. The survey gives the reference share overall, then within five strata: three nominal (occupation, gender, trip purpose) and two ordinal (age class, distance class). Each stratum is scored by a divergence between the simulated and the surveyed shares. Nominal strata take the Jensen--Shannon divergence (JSD) \citep{lin1991divergence} in base 2. Ordinal strata take the Earth Mover's Distance (EMD) \citep{rubner2000earth}, which respects the class order. Both are scaled to $[0, 100]$, the EMD after division by the number of class steps.

One composite number, in points, then adds the overall divergence to the weighted divergence of each stratum. A lower composite means a distribution closer to the survey. A population matching the survey in every stratum would score 0, and spreading every trip uniformly over its options scores about 50.
\[
  \mathcal{C}_{\text{EMD--JSD}} = \mathrm{JSD}^{\text{overall}}
  + \sum_{d\ \text{nominal}} w_d\,\overline{\mathrm{JSD}}^{\,d}
  + \sum_{d\ \text{ordinal}} w_d\,\overline{\mathrm{EMD}}^{\,d}
\]

Each $d$ is a stratum, $w_d$ its weight, and the bar the mean divergence over that stratum. The weights are set by design, not fitted. The overall term has weight 1; age, occupation and purpose 0.5 each; gender and distance 0.3 each (\href{https://hal.science/hal-05797773}{Supplementary Material}: Appendix A).

\begin{figure}
  \centering
  \includegraphics[scale=0.7]{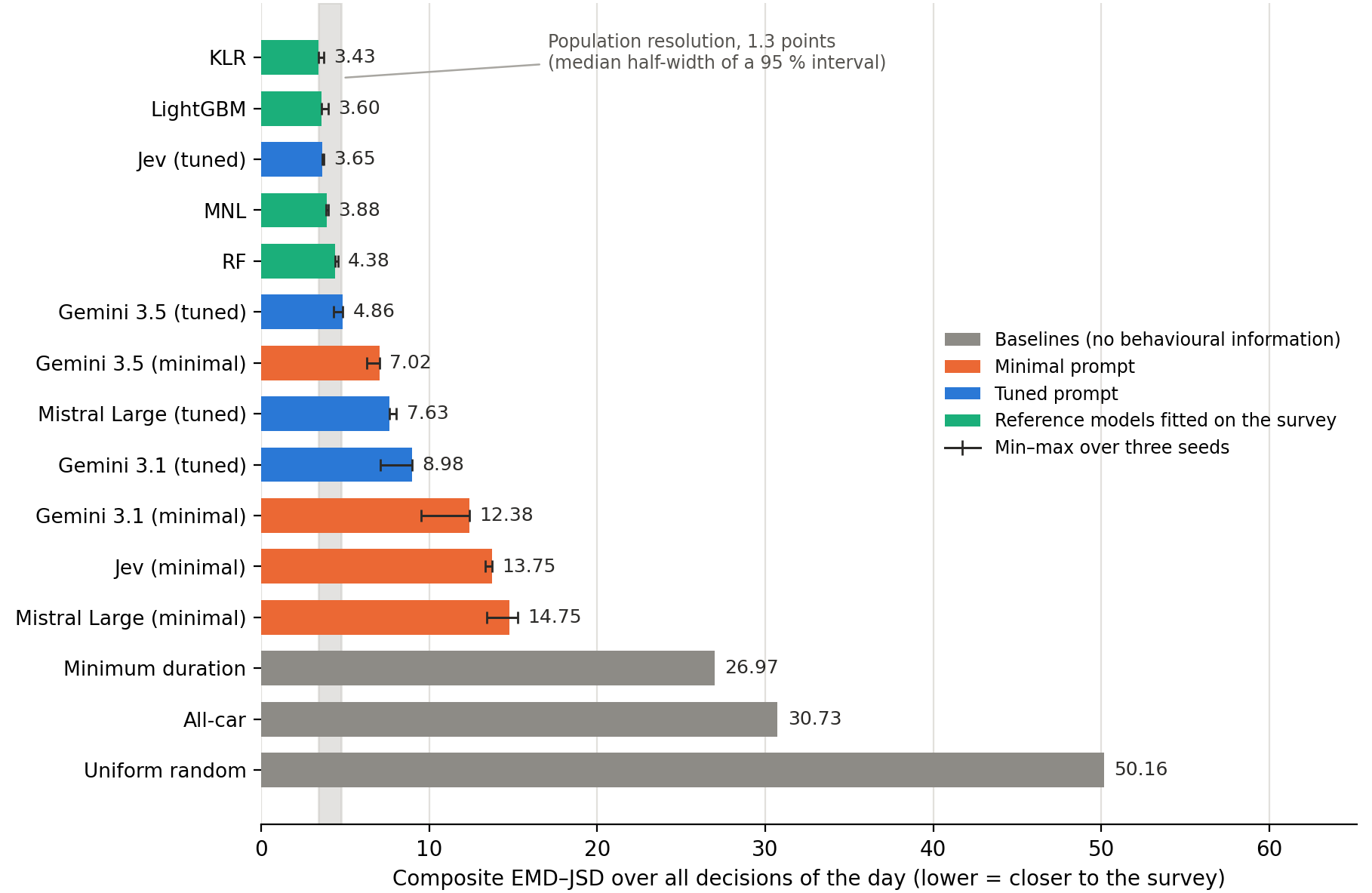}
  \caption{The fifteen decision-makers on the composite. Bars give the seed-42 run, whiskers
  the minimum and maximum over three seeds.}
  \label{fig:scale}
\end{figure}

The best baseline and the reference models set the two ends of the range within which the other decision-makers are read: 26.97 composite points for the former, 3.43 to 4.38 for the latter (Figure \ref{fig:scale}). That end is the level that estimation on the 36 variables of the reference models reaches; a decision-maker can reach or pass it. The second scale compares decisions trip by trip, against the mode each survey respondent reported (Section \ref{sec:individual}).

\paragraph{Repeatability.}
\label{sec:repeat}

Trips by the same person are not independent, so every interval resamples persons rather than trips. The median half-width of a composite's 95\,\% interval over the decision-makers, 1.3 points, measures how much the composite moves when the evaluated persons are resampled. We call it the population resolution and use it as a unit for reading gaps, not as a test of equivalence. Two decision-makers are compared by a paired difference on the persons both scored, with its own 95\,\% interval. The whiskers of Figure \ref{fig:scale} give how far another seed moves each composite (\href{https://hal.science/hal-05797773}{Supplementary Material}: Appendix C).

\paragraph{Release.}
We release the protocol, the code, the seeds, and the evaluation population in the repository\footnote{\url{https://github.com/Ytlse/generative-agents-urban-mobility}}. Our data-use agreement forbids redistributing the survey microdata. The survey is distributed by PROGEDO-ADISP \citep{tisseo2023emc2}.

\subsection{A1: Built-in knowledge of travel}
\label{sec:scale}

Assumption A1 (Section \ref{sec:introduction}) states that GAs bring built-in, albeit partial and approximate, knowledge of people's travel behaviours to their choices. If so, they should beat rules that have none before any tuning. The minimal prompt tests this. Every minimal-prompt condition already has less error than the three baselines (Figure \ref{fig:scale}). The weakest minimal-prompt condition scores 14.75, compared with 26.97 for the best baseline (minimum duration). The models therefore do more than pick the fastest option. Under the minimal prompt, Gemini 3.1, Mistral Large and Jev score between 12 and 15, closing at least half of the distance from the best baseline to the best reference model (3.43). Gemini 3.5, the newer of the two Gemini versions, scores 7.02 and closes 85\,\% of it. A1 therefore holds: the models predict mode choice before any local tuning, with knowledge from their training. What that knowledge consists of, and whether it covers factors that the survey does not record, this comparison cannot tell.

\subsection{A2: Prompt tuning}
\label{sec:tuning}
\label{sec:tuningprocedure}

Assumption A2 (Section \ref{sec:introduction}) states that the behaviour of GAs can be adjusted by tuning the prompt with criteria stated in words. If so, the tuned prompt should lower its error without carrying any number from the survey.

\begin{figure}[h]
  \centering
  \includegraphics[scale=1.3]{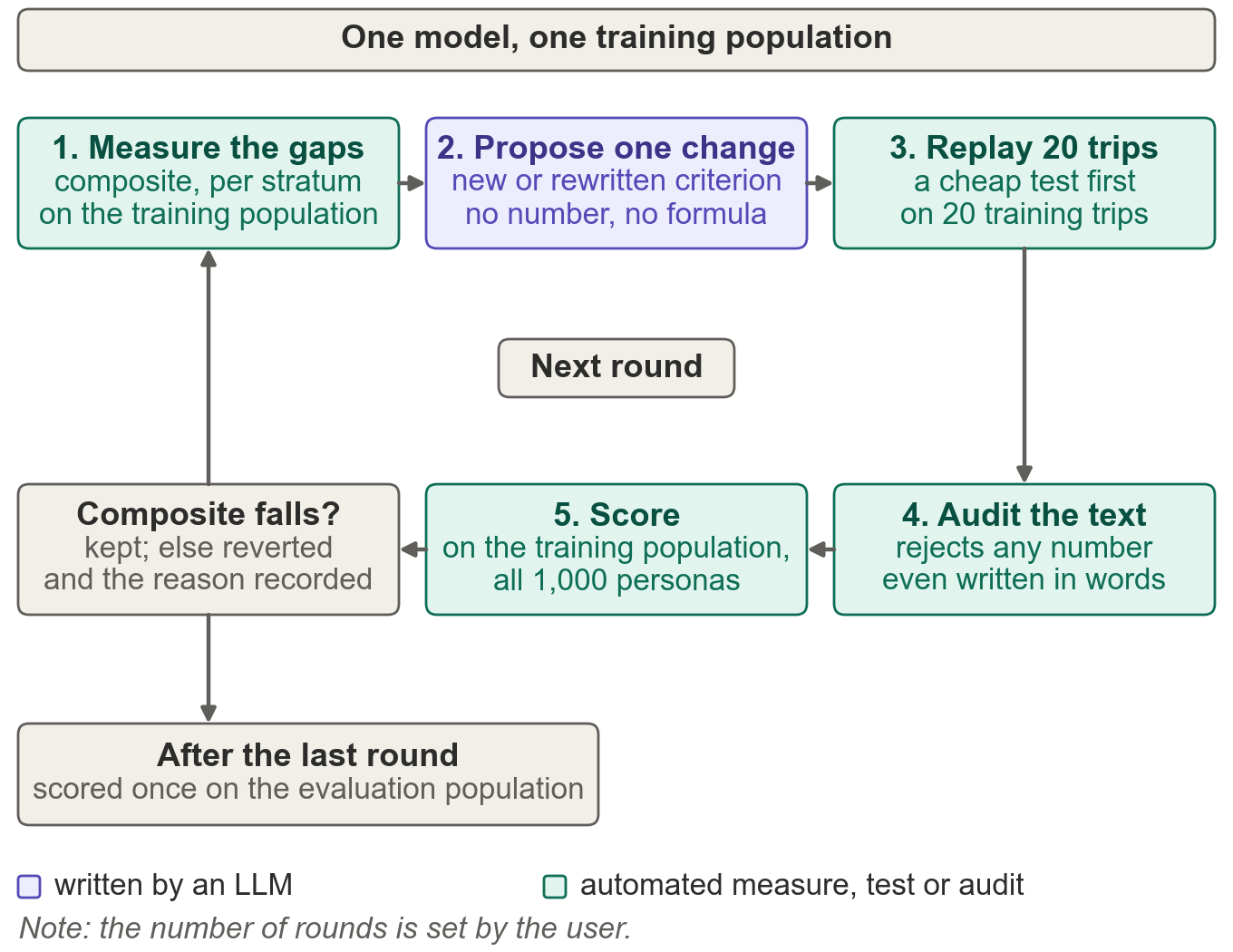}
  \caption{The prompt-tuning loop, run for each model.}
  \label{fig:tuning}
\end{figure}

\paragraph{How the tuned prompt is obtained.} A prompt is tuned for each model. Two models given the same text do not make errors on the same strata, so a prompt that corrects the gaps of one need not suit another. The loop of Figure \ref{fig:tuning} is run automatically for a fixed number of rounds. In each round, an LLM reads the composite per stratum on the training population and proposes one change, a general criterion, new or rewritten, that corrects the reading behind a gap. For example, the model read age as an inability to walk, and so sent older travellers to public transport. 

The resulting criterion reads:
\begin{quote}\itshape
``For elderly or frail people, a continuous, unhurried walk at one's own pace is the natural
mode of independence for short distances, against the strain and stress of public transport
(jolting, risk of falling, steps to climb, standing while waiting with no bench).''
\end{quote}

A criterion describes a situation in plain text and contains no number and no formula. An independent agent, which never writes prompts, audits the whole text and rejects any number, even written in words. The changed prompt is then scored on the 1,000 personas of the training population. It is kept if the composite falls; otherwise the change is rejected. After the last round, the prompt is scored once on the evaluation population. Each model thus ends with its own tuned prompt. That of Gemini 3.5, tuned by Claude Opus 5.5, holds four general criteria on comfort and on constraints such as loads to carry (\href{https://hal.science/hal-05797773}{Supplementary Material}: Appendix B). We call Gemini 3.5 under its tuned prompt the \emph{tuned agent}. It is the best-scoring LLM condition (Figure \ref{fig:scale}).

\begin{figure}
  \centering
  \includegraphics[scale=0.7]{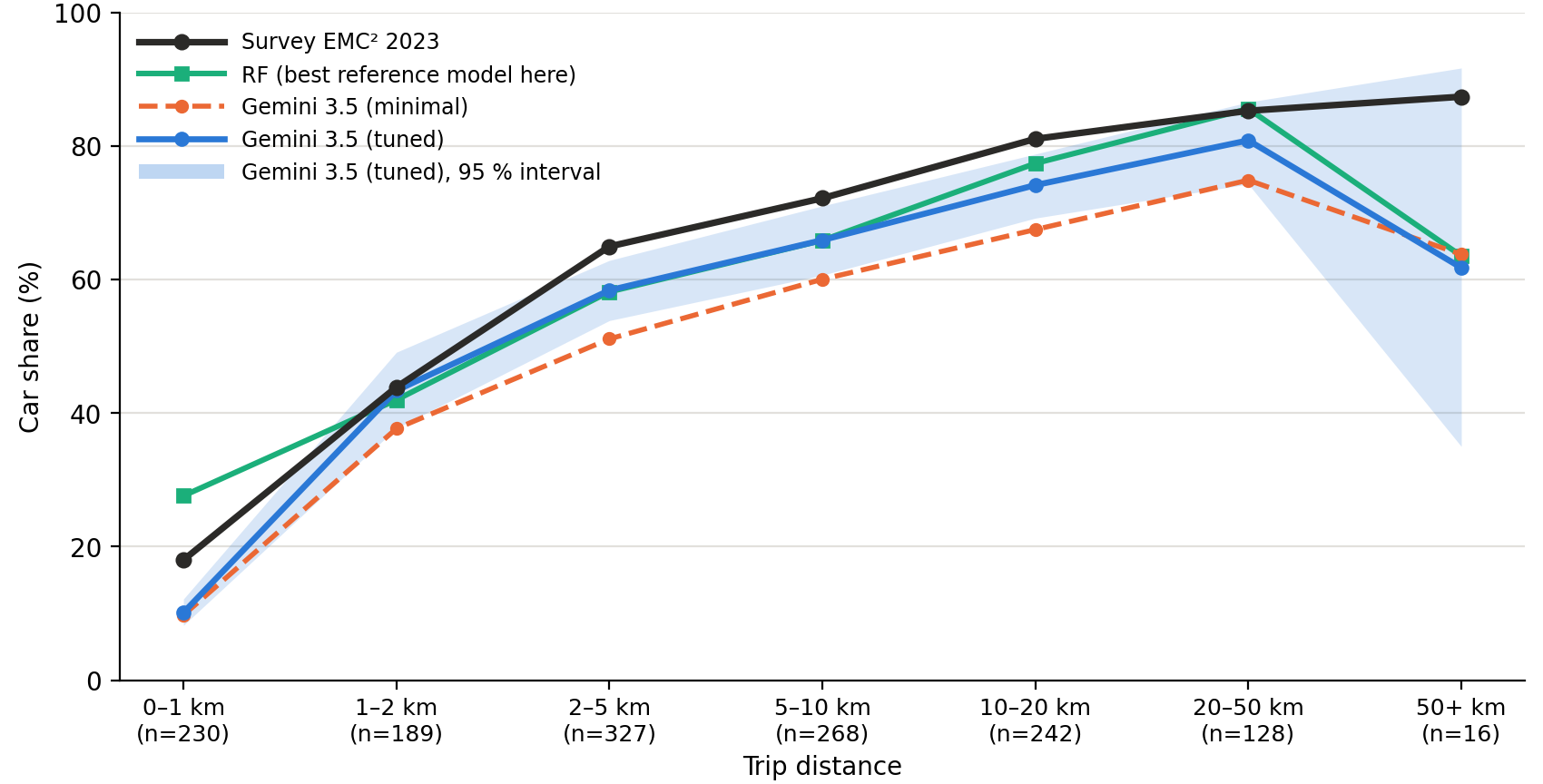}
  \caption{Car share by trip distance (focus on Gemini 3.5).}
  \label{fig:distance}
\end{figure}

\paragraph{Results.} Tuning the prompt significantly lowers the error of all four models (Figure \ref{fig:scale}). The three LLMs gain from 2.16 composite points for Gemini 3.5 (7.02 to 4.86) to 7.12 for Mistral Large (14.75 to 7.63). Jev (tuned prompt) gains 10.10 points (13.75 to 3.65). With Gemini 3.5, tuning brings the car share closer to the survey beyond one kilometre, to about the level of the RF, the reference model closest to the survey across distance bands (Figure \ref{fig:distance}). A gap remains under one kilometre, where the car share stays near 10\,\% before and after tuning, against 18\,\% in the survey.

With the three LLMs tested, tuning moves the public transport share four to eleven points closer to the survey and the bike share by less than one point. A2 therefore holds: criteria stated in words lower the error of all four models, but only in part. The prompt is not independent of the survey, since each wording was kept for bringing the shares closer to it on the training population.

\subsection{A3: Modal split against the reference models}
\label{sec:distributional}

Assumption A3 (Section \ref{sec:introduction}) states that the modal split of GAs comes as close to that of a local population as the split of models fitted on its travel survey. The four reference models set that level, with composites from 3.43 to 4.38 (Figure \ref{fig:scale}). The best LLM condition, the tuned agent, scores 4.86, just above that range, and paired intervals still separate it from three reference models (LightGBM, KLR and the MNL). Measured instead as the L1 error, the sum of the absolute gaps between simulated and surveyed mode shares, the text-based conditions rank in the same order as on the composite and every tuning gain holds (\href{https://hal.science/hal-05797773}{Supplementary Material}: Appendix D). Against the reference models, this measure is stricter: it also separates the tuned agent from the RF, and Jev from LightGBM and KLR. Alignment remains incomplete on some modes and strata. The bike share of the three LLMs stays at 6.8 to 8.1\,\%, whereas the survey records 4.1\,\%, and no decision-maker fits one stratum: the 15--19-year-olds. The survey records 47.4\,\% public transport for them, compared with 25.0\,\% for the RF and 20.8\,\% for the tuned agent.

\label{sec:deliberation}
On the evaluation population, Jev (tuned prompt) scores 3.65, inside the range of the four reference models. No paired composite interval separates it from any of them. Each reference model itself moves by up to one point from the evaluation to the training population. On the training population, on which its prompt was tuned, Jev lies 0.76 points above their range (\href{https://hal.science/hal-05797773}{Supplementary Material}: Appendix C). Our intervals resample persons and do not include this variation, so we do not order decision-makers whose gap is smaller than one point. Under the minimal prompt, however, it scores 13.75, almost seven points behind Gemini 3.5 under the same prompt. A3 therefore holds in part for the LLMs tested: the tuned agent comes close to the range of the reference models without reaching it, and only Jev (tuned prompt) reaches that range.

\subsection{A4: Realism for each individual}
\label{sec:individual}

Assumption A4 (Section \ref{sec:introduction}) states that the choices of GAs are realistic for each individual. If so, an agent that matches the modal split should also choose the mode that each survey respondent reported, trip by trip. The tuned agent and LightGBM, the model with the highest trip-level accuracy (Table \ref{tab:unit}), stand 1.3 composite points apart, a gap equal to the population resolution, and still rank different modes first on 29.6\,\% of trips where both saw at least two modes. Between pairs of reference models, disagreement ranges from 9.1 to 15.2\,\%. Agreement with the survey uses its respondents rather than the evaluation population: 9,621 trips by 2,930 respondents, each with its reported mode. As Rule 1 requires, accuracy is read on the probability mass: it is the probability given to the reported mode, averaged over trips with their survey weights (\href{https://hal.science/hal-05797773}{Supplementary Material}: Appendix A).

On the reported survey days, the tuned agent assigns less mass to the reported mode than all four reference models (Table \ref{tab:unit}). It trails LightGBM by 4.3 accuracy points (95\,\% interval 3.2 to 5.4) and the MNL by 1.3 points (0.2 to 2.4). All four accuracy intervals exclude zero. On cross-entropy, only the interval against LightGBM excludes zero.

\begin{table}
  \caption{Trip-level agreement on the 9,621 reported trips.}
  \label{tab:unit}
  \renewcommand{\arraystretch}{1.3}
  \begin{tabularx}{\textwidth}{@{}>{\centering\arraybackslash}X
                                  >{\centering\arraybackslash}X
                                  >{\centering\arraybackslash}X
                                  >{\centering\arraybackslash}X
                                  >{\centering\arraybackslash}X@{}}
    \hline
    Decision-maker & \multicolumn{1}{c}{Accuracy (\%)} & \multicolumn{1}{c}{Cross-entropy (nats)} & \multicolumn{1}{c}{Bike recall (\%)} & \multicolumn{1}{c}{Walk recall (\%)} \\
    \hline
    LightGBM & 72.0 & 0.298 & 17.8 & 66.1 \\
    KLR & 70.5 & 0.323 & 17.1 & 64.0 \\
    RF & 69.7 & 0.326 & 10.3 & 64.4 \\
    MNL & 69.0 & 0.355 & 17.0 & 61.0 \\
    Minimum duration                                  & 68.1 & n/a   & 25.8 & 19.2 \\
    Gemini 3.5 (tuned) & 67.7 & 0.341 & 21.6 & 48.6 \\
    All-car                                           & 66.7 & n/a   &  2.3 & 16.2 \\
    Gemini 3.5 (minimal) & 64.5 & 0.383 & 22.2 & 44.6 \\
    Jev (tuned) & 64.1 & 0.464 & 17.3 & 57.2 \\
    Gemini 3.1 (tuned) & 58.2 & n/a & 21.1 & 40.8 \\
    Gemini 3.1 (minimal) & 56.9 & n/a & 20.7 & 35.1 \\
    Jev (minimal) & 56.4 & 0.551 & 20.2 & 42.6 \\
    Mistral Large (tuned) & 53.7 & n/a & 19.8 & 55.5 \\
    Mistral Large (minimal) & 49.1 & n/a & 19.7 & 43.1 \\
    Uniform random & 24.3 & n/a & 12.4 & 35.9 \\
    \hline
  \end{tabularx}
\end{table}

\begin{figure}
  \centering
  \includegraphics[scale=0.7]{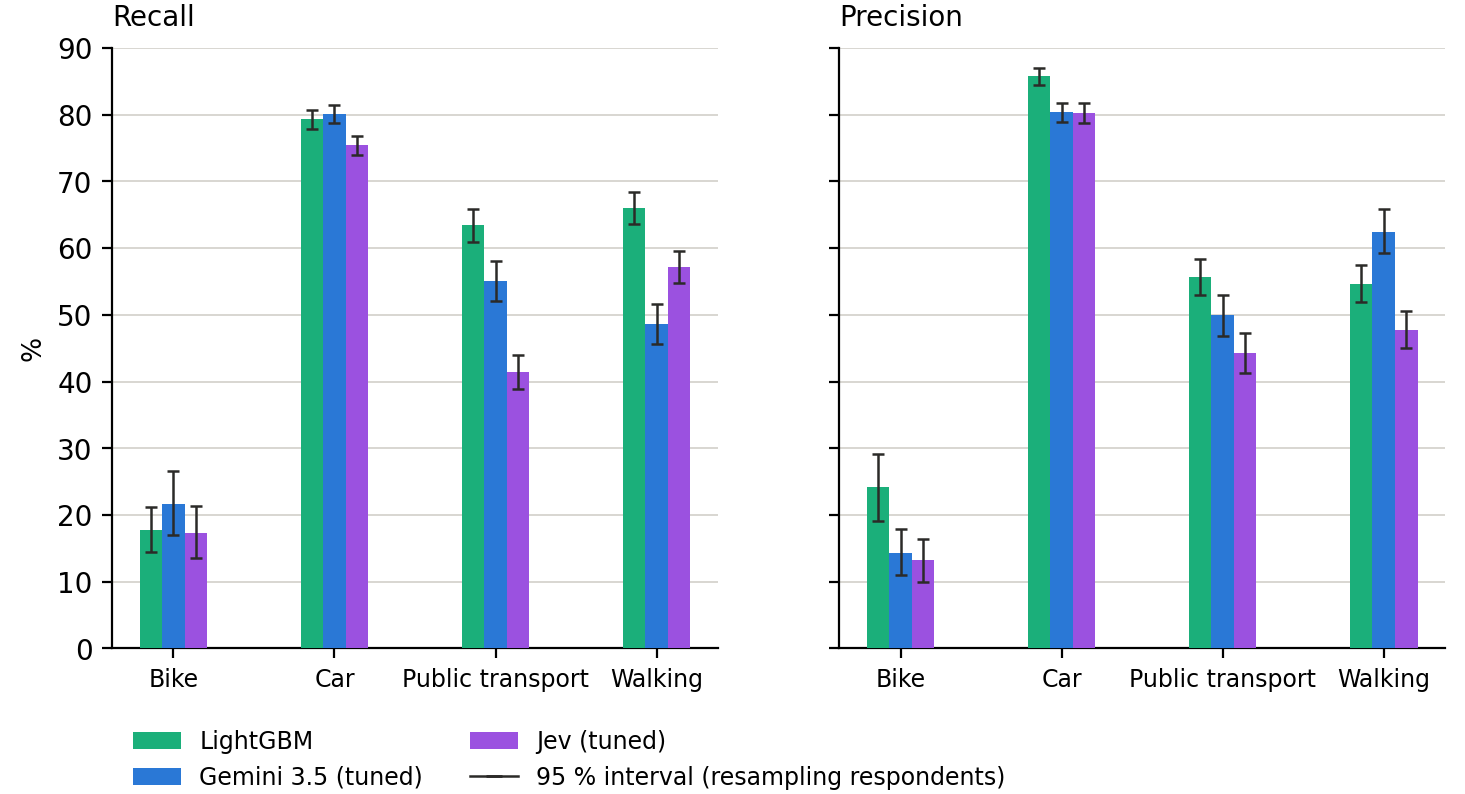}
  \caption{Recall and precision by mode}
  \label{fig:audit-modes}
\end{figure}

\begin{figure}[h]
  \centering
  \includegraphics[scale=1.0]{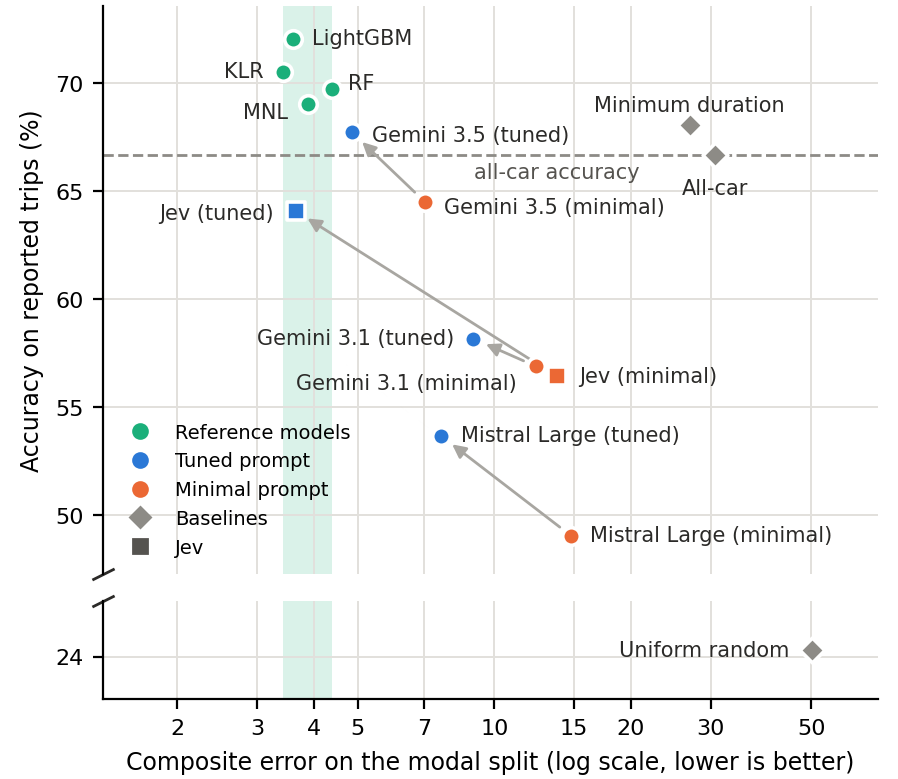}
  \caption{Composite against probability of the reported mode on surveyed trips. Green band, reference models; dashed line, all-car baseline; arrows, minimal to tuned prompt.}
  \label{fig:split-trip}
\end{figure}

On individual accuracy, Jev trails the all-car baseline by 2.6 points (95\,\% interval 1.1 to 4.1), while its composite reaches the range of the reference models (Section \ref{sec:deliberation}). It over-predicts walking at the expense of public transport. Its walk recall reaches 57.2\,\%, compared with about 41\,\% for public transport, and its walk precision falls below that of the two other decision-makers (Figure \ref{fig:audit-modes}). A population can therefore reproduce the modal split of a territory while getting more individual trips wrong than a baseline that takes the car whenever it can (Figure \ref{fig:split-trip}). Assumption A4 therefore does not hold on the ordinary day for the models tested, although the tuned agent's modal split comes close to those of the reference models. Its prompt was tuned on the modal split, not on individual choices.

\section{Sensitivity to unforeseen events}
\label{sec:unforeseen_events}

\label{sec:variables-gap}
A decision-maker restricted to the 36 variables of the reference models cannot condition its choice on an incident or recent news, and these variables offer no measure of such a response. A GA, by contrast, can be told of an event in words. We test in this section how GA reacts compared to the same agent left uninformed. Decisions are made by Gemini 3.1 (tuned prompt) and Gemini 3.5 writes the evening consolidation. The severity of the event sets how long the prompt lists it among recent events (Section \ref{sec:memory}).

\subsection{How the response to an event is checked}
\label{sec:mechanism}

The survey describes a day without incident, and no survey records how residents reacted to this event. No score can therefore say whether a response is right. Three checks of different reach take its place (details in \href{https://hal.science/hal-05797773}{Supplementary Material}: Appendix F).

\emph{Paired control.} The agent is simulated twice over the same weeks, with the same prompt, seeds and settings, once exposed to the event and once as a control without it. Both runs share every model answer until the event first enters a decision prompt.

\emph{Prompts and reasons.} For each decision that changes, we read the model's prompt and the reason it wrote with its answer, to check that the event appears in them. This shows that the event was in view when the choice changed. The reason, however, is text the model generates itself, and can leave out what actually drove the answer \citep{turpin2023language}. It is thus a clue, not a proof.

\emph{Evening questionnaire.} Every simulated evening, apart from any decision, each agent rates the modes out of ten on six items from a mobility survey \citep{adam2025survey}: speed, practicality, comfort, affordability, environment and safety. It also states how much each item weighs in its choice. The environment item serves as a control: the event should not move it.

\subsection{A lived event: the engine noise}
\label{sec:setup}
\label{sec:traced-results}

\begin{figure}[h]
  \centering
  \includegraphics[width=\textwidth]{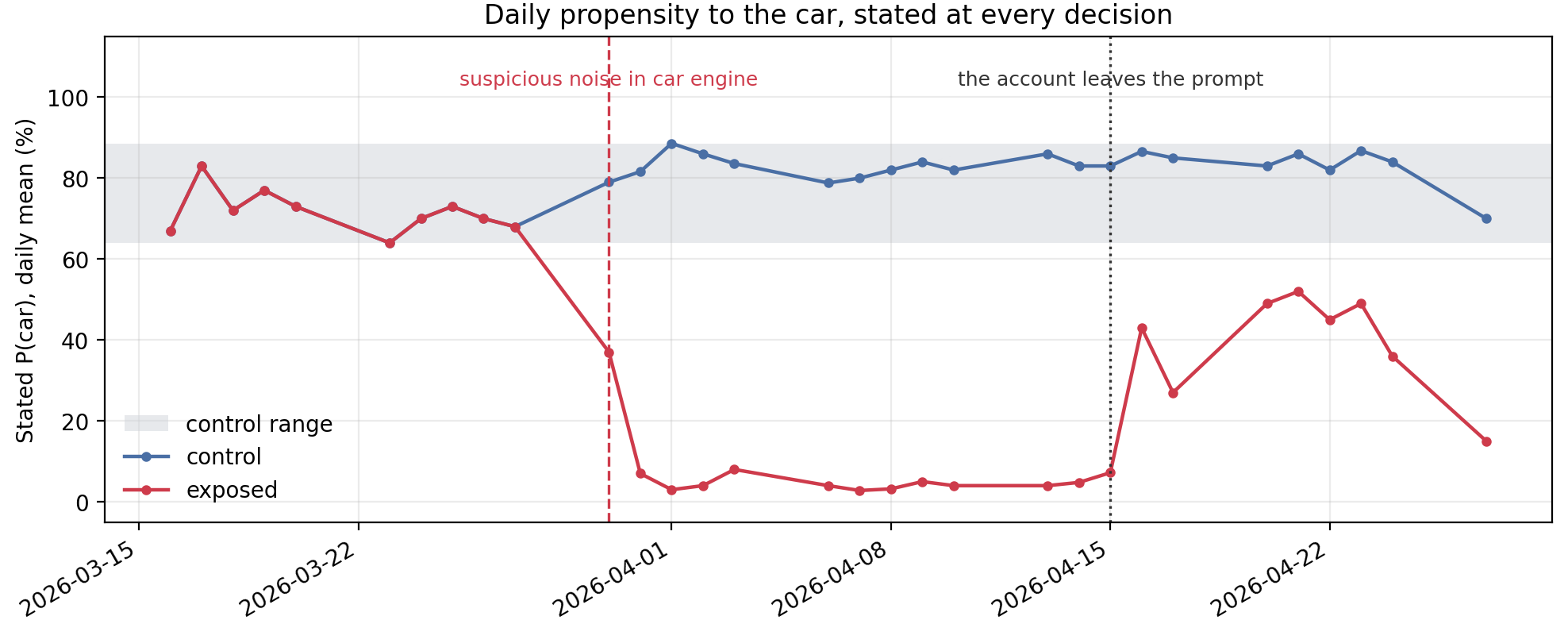}
  \caption{Daily car propensity after the engine noise.}
  \label{fig:propensity}
\end{figure}

The agent is a 58-year-old full-time executive in a two-person household with two cars and no public transport pass. Before the noise, she takes the car on 79\,\% of the decisions where it is offered (41 of 52), as her control does. The noise occurs during a car trip on 30 March, the fifteenth simulated day, and the trip ends 35 minutes late. The car remains offered the next morning. For each trip, we measure the probability that the LLM gives the car. While the noise is in the prompt, the exposed agent never takes the car (0 of 60 offers), whereas her control takes it on 92\,\% of them (56 of 61). Her daily car propensity, the probability the model gives the car averaged over the day's decisions, falls from 64--83\,\% to 3--8\,\% (Figure \ref{fig:propensity}).

On 15 April, the noise leaves the recent-events block. The next day, her propensity rises to 43\,\%. It then stays between 15 and 52\,\% until the run ends twelve days later, below the lowest daily value of her control (64\,\%). Of the offers after 15 April, she takes the car on 36\,\% (14 of 39), against 95\,\% (36 of 38) for her control. The prompts show what carries the effect past expiry. Of the 71 decision prompts after the noise drops out of the recent-events block, 69 recall an evening summary written while it was listed there that names the car's trouble, and so do 20 of the 39 reasons given when the car is offered. Her habits record 16 of 23 morning trips to work on foot. Her 14 car trips without incident, which weaken an event's effect, did not end it within the run.

Stated opinions separate what returns from what lasts. While the noise is in the prompt, the evening questionnaire gives the car lower ratings on four of the six items: safety, practicality, comfort and speed. Safety drops from 7 to 3 out of ten. On 15 April, the first evening without the noise in the prompt, all four ratings are back to their level before the noise or higher. The car's environment rating, which the noise should not affect, does not change. After that, the exposed agent rates walking 4.1 points higher on practicality than before the noise, and the control does not. What lasts is less a distrust of the car than the standing of the alternatives gained while the agent used them.

To sum up, the exposed agent chooses differently from her paired control. The event is present in the prompts when her decisions change, and the environment item stays still. The response outlasts the event's stay in the prompt through the summaries and habits it left in memory. The event is played once, on one agent, and no survey records how residents reacted to it. We therefore claim that this response exists, not how large or how realistic it is.

\section{Conclusion}
\label{sec:conclusion}

We examined the place of text-based decision-makers in generative agents by confronting them with the 2023 household travel survey of Toulouse, France, through five assumptions (A1 to A5). Three simple baselines and four reference models fitted on the survey set the two ends of the range within which their scores are read. Fifteen decision-makers also include text-based models (three LLMs and the typed classifier Jev), each under a minimal and a tuned prompt. The comparison covered population distributions and individual choices, with shared core variables and explicit differences in information and tuning.

We conclude that GAs bring built-in knowledge of travel (A1): even under the minimal prompt, text-based decision-makers already outperform the simple baselines on the ordinary day. Tuning the prompt with criteria stated in words (A2) brings the agent closer to the reference models, with Jev (tuned prompt) reaching the range of the reference models. The modal split of GAs comes close to that of the local population without matching it (A3). Finally, population alignment does not establish individual predictive accuracy (A4). This result makes the joint evaluation of distributions and individual choices essential: a convincing modal split can conceal substantial disagreement about who takes which mode. These findings hold for the Toulouse area, in the condition of the survey. Nevertheless, the methodology could definitely be applied in any case study.

The run with a text-encoded unforeseen event (A5) points out the capability of GAs to adapt to totally new situations, impossible to encode in reference models. That is where GAs may justify their additional cost. Cost further constrains the role of generative models. As an order of magnitude, thirty simulated days of 1{,}000 personas send about 170 million tokens to the LLM, and one measured workload of route decisions costs about \$1 with Jev and about \$50 with Gemini 3.5 (tuned prompt), at the API rates recorded in 2026. The ordinary-day benchmark does not establish an overall fidelity advantage over the best reference models. These results motivate testing hybrid architectures that allocate generative computation to tasks for which it provides a measurable benefit. At city scale, such an architecture would call the GA only when an event occurs, and let a cheaper decision-maker handle the ordinary day.

\section*{Acknowledgements}
This project was carried out as part of the MIDOC Key Challenge (Défi Clé MIDOC), initiated and funded by the Occitanie region. To find out more about the Occitanie Key Challenges: \url{https://www.laregion.fr/Defis-Cles-Occitanie} (in French, accessed October 13, 2026).

\bibliographystyle{ACM-Reference-Format} 
\bibliography{refs}

\end{document}